%Paper: hep-ph/9309252
%From: JIZHI@PHYS.TAMU.EDU
%Date: Thu, 9 Sep 1993 23:39:12 CDT

\magnification 1200
\baselineskip=18pt
\overfullrule=0pt

\input tables.tex

\def\lam{\hbox{$\lambda\kern-6pt^{\_\_}$}}

\def\r{\vbox{\hbox{\raise.5mm\hbox{$>$}}
\kern-18pt\hbox{\lower1.5mm\hbox{$\sim$}}}}
\def\l{\vbox{\hbox{\raise.5mm\hbox{$<$}}
\kern-18pt\hbox{\lower1.5mm\hbox{$\sim$}}}}

\def\at{\hbox{$^{\ast}$}}
\def\leaderfill{\leaders\hbox to 1em{\hss.\hss}\hfil}
\def\sqr#1#2{{\vcenter{\vbox{\hrule height.#2pt
        \hbox{\vrule width.#2pt height#1pt \kern#1pt
          \vrule width.#2pt}
        \hrule height.#2pt}}}}

\line{\hfil CTP-TAMU-32/93}
\line{\hfil NUB-TH-3066/93}
\line{\hfil SSCL-Preprint-440}
\bigskip
\centerline{\bf TESTING SUPERGRAVITY GRAND UNIFICATION AT FUTURE}
\centerline{\bf ACCELERATOR AND UNDERGROUND EXPERIMENTS}
\bigskip
\centerline{R. Arnowitt$^{a)}$ and Pran Nath$^{b)}$}
\centerline{{\footnote\at{Permanent address}}$^a$Center for Theoretical
Physics, Department of Physics}
\centerline{Texas A\&M University, College Station, TX  77843-4242}
\centerline{$^a$Physics Research Divsion, Superconducting Super}
\centerline{Collider Laboratory, Dallas, TX  75237}
\centerline{$^b$Department of Physics, Northeastern University}
\centerline{Boston, MA  02115}
\bigskip
\centerline{ABSTRACT}
\medskip

The full parameter space of supergravity grand unified theory with $SU(5)$ type
$p \rightarrow \bar{\nu} K$ proton decay is analysed using renormalization
group induced electroweak symmetry breaking under the restrictions that the
universal scalar mass $m_o$ and gluino mass are $\leq 1$ TeV (no extreme fine
tuning) and the Higgs triplet mass obeys $M_{H_3}/M_G < 10$.  Future proton
decay experiments at SuperKamiokande or ICARUS can reach a sensitivity for the
$\bar{\nu} K$ mode of $(2-5) \times 10^{33}$ yr allowing a number of
predictions concerning the SUSY mass spectrum.  Thus either the $p \rightarrow
\bar{\nu} K$ decay mode will be seen at these experiments or a chargino of mass
$m_{\tilde{W}} < 100$ GeV will exist and hence be observable at LEP2.  Further,
if $(p \rightarrow \bar{\nu} K) > 1.5 \times 10^{33}$ yr, then either the light
Higgs has mass $m_h \leq 95$ GeV or $m_{\tilde{W}} \leq 100$ GeV i.e. either
the light Higgs or the light chargino (or both) would be observable at LEP2.
Thus, the combination of future accelerator and future underground experiments
allow for strong experimental tests of this theory.
\vfill\eject

\noindent
1.  INTRODUCTION
\medskip

The observation that, for a supersymmetric mass spectrum, the three coupling
constants of the Standard Model measured at the scale $Q = M_Z$, $\alpha_1
(M_Z) \equiv (5/3) \alpha_Y (M_Z)$, $\alpha_2 (M_Z)$~and~$\alpha_3 (M_Z)$,
unify at the Gut scale $Q = M_G$ to a common value $\alpha_G$ [1] has lead to
considerable effort to deduce additional consequences of supergravity grand
unification models [2,3].  Unification takes place at $M_G \sim 10^{16}$ GeV
with the SUSY particles having mass at $M_S \sim 10^{2.5}$ GeV provided only
one pair (the minimal number) of Higgs doublets exist.  It is possible that the
fact that the three coupling constants meet at a point at $10^{16}$ GeV is
merely a numerical accident without significance.  If, however, one accepts
this result as a guide to new physics, it suggests first the validity of grand
unification, and second that the particle spectrum above the electroweak scale
and up to the Gut scale is that of the supersymmetrized Standard Model with
only one pair of Higgs doublets.  (More pairs of Higgs doublets leads to too
small a value of $M_G$ and hence too rapid $p \rightarrow e^+ \pi^o$ proton
decay.)

It is of interest that the unification of the couplings is not a property of
low energy supersymmetry alone.  Thus the 5/3 factor relating $\alpha_1$ to
$\alpha_Y$ is needed to achieve unification, and this reflects on how
$\alpha_Y$ is embedded into a grand unification group $G$ at the Gut scale.
(Examples of acceptable choices of $G$ are $SU(5)$, $O(10)$, $E_6$ etc. but not
$[SU(3)]^3$).  There is no reason in low energy SUSY theory to insert the 5/3
factor, and hence no reason to expect coupling constant unification from purely
low energy considerations.  Further, supersymmetry must break spontaneously,
and no phenomenologically acceptable way of doing this in low energy global
supersymmetry has been constructed.  In supergravity, however, spontaneous
symmetry breaking of supersymmetry in the ``hidden'' sector occurs naturally,
either at the tree level [4] or via condensates [5].

For these reasons, activity over the past two years has centered around
obtaining predictions of different supergravity grand unified models [6-15].
In these models, our lack of knowledge of the physics of the hidden sector can
be parameterized in terms of four parameters:  $m_o$ (universal scalar mass),
$m_{1/2}$ (universal gaugino mass), $A_o$ (cubic soft breaking parameter) and
$B_o$ (quadratic soft breaking parameter).  Refs. [6,7] treat the general
$SU(5)$ supergravity, while Ref. [10] examines this in the special case where
$B_o = A_o - m_o$.  The constaints of proton decay $p \rightarrow \bar{\nu} +
K^+$ are included in Refs. [6,7].  The No-Scale models ($m_o = A_o = 0$ and
also $B_o = 0$) are examined without proton decay (as would be the case for the
flipped $SU(5)$ model [16]) in Refs. [11,14] and with the constaint of proton
decay in Ref. [12].  Cosmological constraints are discussed in Ref. [13].
Models with $O (10)$ symmetry were examined in Ref. [15].

In this paper we will consider a general class of supergravity Gut models
(defined explicitly in Sec. 2 below) which allow for an ``$SU(5)$-type'' proton
decay in the $p \rightarrow \bar{\nu} + K^+$ mode.  [As discussed in Sec. 2,
it is difficult to prevent this type of proton decay for $SU(5)$-type models,
except for the case of flipped $SU(5)$.]  Present proton decay data [17]
significantly restricts the parameter space.  Thus in previous work it was
shown that it leads generally to a lower bound on $m_o$ and an upper bound on
the gluino mass $m_{\tilde{g}} = (\alpha_3/\alpha_G) m_{1/2}$, with both
squarks and gluino probably requiring the SSC or LHC to be seen [6,7].  Since
$m_o$ is large, radiative breaking [18] of $SU(2) \times U(1)$ at the
electroweak scale generally implies that the $\mu$ parameter (which scales the
coupling of the two Higgs  doublets $H_1$~and~$H_2$ in the superpotential)
obeys $\mu^2 >> M_Z^2$.  This leads to a number of scaling laws between the
charginos ($\tilde{W}_i, i = 1,2$), neutralinos ($\tilde{Z}_i, i = 1 \cdots 4$)
and the gluino:

$$\eqalignno{2 m_{\tilde{Z}_1}&\cong m_{\tilde{W}_1} \cong
m_{\tilde{Z}_2}&(1.1a)\cr
m_{\tilde{Z}_{3,4}}&\cong m_{\tilde{W}_2} >> m_{\tilde{Z}_1}&(1.1b)\cr
m_{\tilde{W}_1}&\simeq {1\over 3}~m_{\tilde{g}} (\mu < 0);~~m_{\tilde{W}_1}
\simeq {1\over 4}~m_{\tilde{g}} (\mu > 0)&(1.1c)\cr}$$

\noindent
(where $m_{\tilde{Z}_i} < m_{\tilde{Z}_j}$~for~$i < j$ etc.)  Bounds on the
Higgs masses ($h, H = CP$ even states, $A = CP$ odd and $H^{\pm} =$~charged
Higgs) also are obtained [6,7]:

$$m_h~\l~110 GeV;~~m_A \cong m_H \cong m_{H^{\pm}} >> m_h\eqno(1.2)$$

\noindent
Further, there arises an upper bound on the top quark mass, $m_t~\l~175$ GeV
with the first two generations of squarks and all three generations of sleptons
approximately degenerate.  The third generation of squarks are highly split.
Finally we mention that bounds exist on $\tan \beta~\equiv <H_2>/<H_1>$ of
[6] $\tan \beta~\l~7$ and on $A_t$ (top quark $A$ parameter at the
electroweak scale) of $\mid A_t \mid~\l~1.5$.

In this paper we examine additional constraints that can be expected to arise
from future proton decay experiments such as SuperKamiokande and ICARUS, and
from future  accelerator experiments at LEP200 and the Tevatron.  We will see
that, with the expected sensitivities, when one {\it combines} the results of
underground and accelerator experiments one can obtain strong tests of this
class of models.  While each type of experiment by itself can limit the allowed
parameter space, together they can test the validity of supergravity
models with proton decay.
\bigskip

\noindent
2.  REVIEW OF FORMALISM
\medskip

We summarize briefly here the formalism used in calculating consequences of
supergravity Gut models.  The class of supergravity Gut models we will consider
are defined by the following assumptions:

\item{(i)} There exists a hidden sector which is gauge singlet with respect to
the physical sector gauge group $G$ which breaks supersymmetry.  This breaking
communicates to the physical sector only gravitationally.  [Thus in the
super Higgs mechanism, this condition is realized by an additive superpotential
$W = W_{{\rm phys.}} (z_a) + W_{{\rm hidden}} (z)$ where supersymmetry is
spontaneously broken by the VEVs $<z> = O (\kappa^{-1})$, $\kappa^{-1} =
M_{P\ell} \equiv (\hbar c/8\pi G_N)^{1/2} (M_{P\ell} = 2.4 \times 10^{18}$
GeV).
Here, the $\{z_a\}$ are the physical fields.]

\item{(ii)} A Gut sector exists which breaks $G$ to $SU(3)_C \times SU(2)_L
\times U(1)_Y$ at scale $M_G$.

\item{(iii)} After integrating out the super heavy fields (and eliminating the
super Higgs fields) the only light particles remaining below $M_G$ are the
supersymmetric Standard Model particles with one pair of light Higgs doublets.

\item{(iv)} The super Higgs couplings in the Kahler potential are generation
blind.

\noindent
Conditions (ii) and (iii) are what is implied by the analysis of the coupling
constant  unification.  Condition (i) is needed to maintain the gauge hierarchy
and (i) and (iv) together guarantee the suppression flavor changing neutral
interactions.

Conditions (i)-(iv), plus the requirement that the gauge kinetic function
$f_{\alpha\beta}$ and Kahler metric $d_j^i$ can be expanded in a series scaled
by $\kappa~(f_{\alpha\beta} = c_{\alpha\beta} + \kappa c_{\alpha\beta i} z^i +
\cdots$, $d_j^i = c_j^i + \kappa c_{jk}^i z^k + \cdots$, $\{z_i\} = \{z_a -
<z_a>$, $z - <z>\}$) then leads to the following general theorem [19]:  The
renormalizable interactions (arising equivalently from the $\kappa \rightarrow
0$ limit) of a general model is characterized at $M_G$ by an effective
superpotential with quadratic and cubic terms $W = W^{(2)} + W^{(3)}$ given by

$$W = \mu_o H_1 H_2 + [\lambda_{ij}^{(u)} q_i H_2 u_j^C + \lambda_{ij}^{(d)}
q_i H_1 d_j^C + \lambda_{ij}^{(e)} \ell_i H_1 e_j^C],\eqno(2.1a)$$

\noindent
an effective potential given by

$$V = \{\sum\limits_a \mid {\partial W\over \partial z_a} \mid^2 + V_D\} +
[m_o^2 \sum\limits_a z_a z_a^{\ast} + (A_o W^{(3)} + B_o W^{(2)} +
h.c.)]\eqno(2.1b)$$

\noindent
and a universal gaugino mass term ${\cal L}_{{\rm mass}}^{\lambda} = - m_{1/2}
\bar{\lambda}^{\alpha} \lambda^{\alpha}$.  In Eq. (2.1), $q_i, l_i, H_1, H_2$
are $SU(2)_L$ quark, lepton and Higgs doublets ($i = 1,2,3$ is the generation
index), $u_i^C, d_i^C, e_i^C$ are conjugate singlets, $V_D$ is the usual $D$
term, $\lambda_{ij}^{(u,d,e)}$ are the usual Yukawa coupling constants and
$\{z_a\}$ now represents the scalar components of the light chiral multiplets.
Thus aside from the Yukawa coupling constants of the Standard Model, the theory
depends upon four soft breaking parameters $m_o, m_{1/2}, A_o, B_o$ (which
parameterize the properties of the hidden sector) and the parameter $\mu_o$.

The above discussion constructs the renormalizable interactions valid below
$M_G$.  Since $M_G$ is close to $M_{P\ell}$, i.e. $M_G/M_{P\ell} \approx
10^{-2}$, one may suspect the existance of additional ``Planck slop'' terms.
Since the nature of these are unknown, we omit them in the following
discussions.  However, their possible existance implies that the models we are
considering may have errors of order of a few percent.

Supergravity Gut models offer a natural origin of electroweak breaking.  Thus
from Eq. (2.1b), all spin zero particle have a (mass)$^2$ of $m_o^2 > 0$ at
the Gut scale.  Running the renormalization group equations (RGE) down to the
electroweak scale, one finds that the $H_2$ (mass)$^2$ can turn negative
triggering electroweak breaking [18].  The Higgs part of the effective
potential is

$$\eqalign{V_H&= m_1^2 (t) \mid H_1 \mid^2 + m_2^2 (t) \mid H_2 \mid^2 - m_3
(t)^2 (H_1 H_2 + h.c.)\cr
&+{1\over 8}~[g_2^2 (t) + g_Y^2 (t)] [\mid H_1 \mid^2 - \mid H_2 \mid^2]^2 +
\Delta V_1\cr}\eqno(2.2)$$

\noindent
where $t = ln [M_G^2/Q^2]$ is the running parameter, $m_i^2 (t) =
m_{H_i}^2 (t) + \mu^2 (t)$, $i = 1,2$, $m_3^2 (t) = - B (t) \mu
(t)$~and~$\Delta V_1$ is the one loop correction [20].  At $Q = M_G$ (i.e. $t =
0$) the running masses then obey the boundary conditions $m_i^2 (0) =
m_o^2 + \mu_o^2$, $m_3^2 (0) = - B_o \mu_o$~and~$g_2^2 (0) = (5/3)
g_Y^2 (0) = 4\pi \alpha_G$.  Minimizing $V_H$ with respect to $v_i
\equiv <H_i^o>$, $i = 1,2$ yields the equations:

$${1\over 2}~M_Z^2 = {\mu_1^2 - \mu_2^2 \tan^2 \beta\over \tan^2 \beta -
1}~;~\sin 2\beta = {2 m_3^2\over \mu_1^2 + \mu_2^2}\eqno(2.3)$$

\noindent
where $\mu_i^2 = m_i^2 + \Sigma_i$~and~$\Sigma_i$ are the loop corrections:

$$\Sigma_i = \Sigma_a (-1)^{2s_a} n_a [M_a (v_i)]^2 ln [M_a^2/\sqrt{e} Q^2]
(\partial M_a^2/\partial v_i)\eqno(2.4)$$

\noindent
($M_a, s_a, n_a$ are the mass, spin, and number of helicity states of particle
$a$.)  In practice, Eqs. (2.3) is insensitive to the value of $Q$ in the
electroweak scale [21] and one may set $Q = M_Z$.  (Also, for most of the
parameter space, the loop corrections are small.)

The RGE allow the parameters in Eqs. (2.3) to be expressed in terms of the Gut
scale parameters of Eqs. (2.1).  It is convenient to use Eqs. (2.3) to
eliminate  $\mu_o^2$~and~$B_o$ in terms of $\tan \beta$ and the other Gut
parameters.  Thus one is left with

$$m_o, m_{1/2}, A_o, \tan \beta~{\rm and}~m_t\eqno(2.5)$$

\noindent
as unknown constants since $M_G$~and~$\alpha_G$ are determined by the
unification analysis.  (Using two loop RGE and neglecting all thresholds we
find $M_G = 10^{16.19 \pm 0.34}$ GeV, $\alpha_G^{-1} = 25.7 \pm 1.7$ and the
common SUSY particle mass is $M_S = 10^{2.37 \pm 1.0}$ GeV, where the error is
due to the uncertainty in $\alpha_3$ which we take as $\alpha_3 (M_Z) = 0.118
\pm 0.007$ [22].)  Since the sign of $\mu_o$ is not determined by Eq. (2.3)
there are two branches:  $\mu > 0$~and~$\mu < 0$.

If one specifies the five parameters of Eq. (2.5) one may explicitly calculate
the masses of all 32 SUSY particles  (12 squarks, 9 sleptons, 1 gluino, 2
charginos,  4 neutralinos and 4 Higgs bosons).  A characteristic example is
given in Fig. 1.  Note that the first two generations of squarks and all three
generations of sleptons are nearly degenerate.  However, the third   generation
of squarks is widely split, a feature that needs to be taken into account in
phenomenological analyses.  The charginos  and neutralinos exhibit the scaling
of Eqs. (1.1) and the Higgs bosons the relations of Eq. (1.2).  In general,
there are 27 predictions available among the 32 SUSY masses (28 once the top
mass is known), so the theory a priori has a great deal of predictive power.

All supergravity Gut models predict proton decay in the mode $p \rightarrow e^+
+ \pi^o$, and most possess the SUSY mode $p \rightarrow \bar{\nu} + K^+$.  We
consider here models with $SU(5)$-type proton decay defined by the following:

\item{(i)} The Gut group $G$ contains an $SU(5)$ subgroup [or is $SU(5)$].

\item{(ii)} The matter that remains light after $G$ breaks to $SU(3)_C \times
SU(2)_L \times U(1)_Y$ is embedded in the usual way in the $10 + \bar{5}$
representations of the $SU(5)$ subgroup.

\item{(iii)} After $G$ breaks, there are only two light  Higgs doublets which
interact with matter, and these are embedded in the 5 and $\bar{5}$ of the
$SU(5)$ subgroup.

\item{(iv)} There is no discrete symmetry or condition that forbids the proton
decay amplitude.

\noindent
Under the above conditions (which can arise in a number models e.g. $G =
SU(5)$, $O (10)$, $E_6$ etc.) there is a model independent amplitude for the $p
\rightarrow \bar{\nu} + K^+$ decay arising from the exchange of the superheavy
Higgsino color triplet of mass $M_{H_3}$ [23,24].  A characteristic diagram is
shown in Fig. 2.  (Diagrams with other gauginos can also enter, though these
contributions are generally quite small.)

The total decay rate is $\Gamma (p \rightarrow \bar{\nu} K) = \Sigma_i \Gamma
(p \rightarrow \nu_i K)$ where $i = e, \mu, \tau$.  The CKM matrix elements
enter at vertices in Fig. 2 allowing all three generations to enter in the loop
integral.  Thus one may write

$$\Gamma (p \rightarrow \bar{\nu} K) = {\rm Const} (\beta_p/M_{H_3})
\sum\limits_i \mid B_i \mid^2\eqno(2.6)$$

\noindent
where $B_i$ is the amplitude of the $\bar{\nu}_i K$ mode.  $\beta_p$ is given
by

$$\beta_p U_L^{\delta} = \varepsilon_{abc} \varepsilon_{\alpha\beta} < 0 \mid
d_{aL}^{\alpha} u_{bL}^{\beta} u_{cL}^{\delta} \mid p>\eqno(2.7)$$

\noindent
where $U_L^{\delta}$ is the proton wave function.  Lattice gauge calculations
give [26] $\beta_p = (5.6 \pm 0.8) \times 10^{-3}$ GeV$^{-1}$.  In general the
first generation contributions to Eq. (2.6) are negligible and may be
neglected.  One has then [24]

$$B_i = {m_i^d V_{i1}^+\over m_s V_{21}^+}~[P_2 B_{i2} + {m_t V_{31}
V_{32}\over m_c V_{21} V_{22}}~P_3 B_{i3}] {L\over \sin 2\beta}\eqno(2.8)$$

\noindent
where $B_{ia}$ is the loop amplitude when generation a squarks (or sleptons)
enter in the loop, $V_{ij}$ is the CKM matrix and $m_i^d, m_s$, etc. are quark
masses.  The $P_{\dot a}$ are additional CP violating phases arising in the
dimension 5 operators.  To minimize the constraints imposed by proton decay, we
will assume in the following that $P_2/P_3 = -1$, i.e. second and third
generation contributions distructively interfere.  Detailed formulae for
$B_{ia}$ are given in Ref. [24].

Proton decay of this type is a characteristic feature of supergravity grand
unification and one must do special things to avoid it.  Thus the flipped
$SU(5) \times U(1)$ model [16] suppresses the $p \rightarrow \bar{\nu} + K^+$
mode by violating condition (ii) i.e. flipping the embedding of the quarks and
leptons.  Models that impose discrete symmetries to prevent this mode generally
have more than one pair of light Higgs doublets, and sometimes relatively light
Higgs color triplets [25].  This can produce problems with the unification of
the coupling constants.  One may construct models which fine tune the proton
decay amplitude to zero.  Thus consider a model with an arbitrary number of
superheavy color Higgs triplets $H_i$, $\bar{H}_i$ and chose the basis where
only $H_1$, $\bar{H}_1$ couple to matter:

$$\bar{H}_1 J + \bar{K} H_1 + M_{ij} \bar{H}_i H_j\eqno(2.8)$$

\noindent
Here $J$, $\bar{K}$ are bilinear matter sources and $M_{ij}$ is the superheavy
Higgs mass matrix.  Eliminating the superheavy fields, the proton decay
amplitude is then $- \bar{K} (M^{-1})_{11} J$, and if one five tunes the
mass matrix so that $(M^{-1})_{11} = 0$, proton decay is suppressed.
One must also arrange the Gut sector of the model so that only two Higgs
doublets remain light.  While it is possible to construct such models, we will
not pursue them here as they are somewhat artificial.
\bigskip

\noindent
3.  CONSTRAINTS FROM FUTURE EXPERIMENTS
\medskip

We examine now the constraints that can be obtained from proton decay and
collider experiments.  We allow the five parameters, $m_o$, $m_{1/2}$, $A_o$,
$\tan \beta$~and~$m_t$ to range over the entire parameter space subject to:
(i) there be no violation of current experimental bounds on the particle
masses,
(ii) radiative breaking of $SU(2) \times U(1)$ take place,  (iii) proton decay
constraints be obeyed, and (iv) $m_o$, $m_{\tilde{g}} < 1$ TeV (no extreme fine
tuning) and $M_{H_3}/M_G < 10$ (where $M_G \simeq 1.5 \times 10^{16}$ GeV is
grand unification mass neglecting heavy particle thresholds).  The last
condition takes into account the splitting that can arise in the superheavy
particle spectrum.  The upper bound on $M_{H_3}$ ($\simeq 2 \times 10^{17}$
GeV) is as large as is reasonable to assume and still not have to worry about
Planck slop terms.  (It is also what arises naturally in simple models [27] of
the Gut sector.)

The current 90\% C.L. experimental bounds are [17] $\tau (p \rightarrow e^+
\pi^o) > 5.5 \times 10^{32}$ yr and $\tau (p \rightarrow \bar{\nu} K^+) > 1
\times 10^{32}$ yr.  For SUSY theories one expects [28] $\tau (p \rightarrow
e^+ \pi^o) \sim 10^{31 \pm 1}$ ($M_V/6 \times 10^{14}$ GeV)$^4$ yr where $M_V$
is the superheavy vector boson mass.  Super Kamiokande should be sensitive to
the $e^+ \pi^o$ mode up to a lifetime of $1 \times 10^{34}$ yr and up to
$2 \times 10^{33}$ yr for the $\bar{\nu} K$ mode [29], while ICARUS expects to
be
sensitive to the $\bar{\nu} K$ mode up to a lifetime of $5 \times 10^{33}$ yr
[30].  For Super Kamiokande to see the $e^+ \pi^o$ decay mode would require
$M_V~\l~6 \times 10^{15}$ GeV.  On the other hand, the current Kamiokande data
for $p \rightarrow \bar{\nu} K$ requires $M_{H_3}~\r~(0.8) M_G \simeq 1.2
\times
10^{16}$ GeV [6,7].  In simple models [27], one requires $M_{H_3}~\l~3 M_V$, in
order that the Gut physics remain treatable by perturbation theory.  Thus
it is not too likely that Super Kamiokande will be able to see the $p
\rightarrow e^+ \pi^o$ mode, and if it were observable there, the $p
\rightarrow
\bar{\nu} K$ decay would expected to be very copious.

We turn now to consider the $p \rightarrow \bar{\nu} K$ decay in detail.  For
$M_{H_3}/M_G = 3$ it was seen that current Kamiokande data requires [6]
$m_o~\r~500$ GeV, $m_{\tilde{g}}~\l~450$ GeV, $1.1 < \tan \beta < 5$, and
$\mid A_t \mid~\l~1.2$.  One can understand this qualitatively from the fact
that in the limit of large $m_o$, the amplitude for the second generation
contribution to $B_2$ is given approximately by $B_2 \simeq - 2
(\alpha_2/\alpha_3 \sin 2\beta) m_{\tilde{g}}/m_{\tilde{q}}^2$ where the
squark mass is $m_{\tilde{q}}^2 \simeq m_o^2 + a m_{\tilde{g}}^2$, $a \simeq
0.65$.  Thus the proton decay constraint favors a heavy squark, a lighter
gluino and a small $\tan \beta$.  As $M_{H_3}/M_G$ is increased, the lower
bound on $m_o$ decreases and the upper bound on $m_{\tilde{g}}$ increases so
that at $M_{H_3}/M_G~\r~7$, the current data can be satisfied for
$m_{\tilde{g}}~\l~1$ TeV (i.e. for $m_o$ small, $B_2$ again decreases for very
large $m_{\tilde{g}}$:  $B_2 \sim 1/m_{\tilde{g}}$).  The bands on $\tan
\beta$~and~$\mid A_t \mid$ also increase somewhat.

To see the reach of future proton decay experiments, we consider a fixed value
of $m_o$~and~$m_t$, and calculate the maximum lifetime $\tau (p \rightarrow
\bar{\nu} K)$ as all other parameters are varied over the entire allowed
parameter space.  This is shown in Figs. (3a,b,c) for $m_t = 125$ GeV, 150 GeV
and 170 GeV $(\mu < 0)$ for the three values $M_{H_3}/M_G = 3, 6$, and 10.
(The $\mu > 0$ lifetimes are slightly shorter, but show a similar behavior.)
One sees that the entire parameter domain of $m_o < 1000$ GeV will be
accessible to ICARUS for $M_{H_3}/M_G < 6$ (and accessible to Super Kamiokande
for $m_o~\l~800$ GeV).  Thus if $M_{H_3}/M_G < 6$, proton decay should be seen
at ICARUS for this class of models.  Fig. 4 shows the maximum value of $\tau
(p \rightarrow \bar{\nu} K)$ for $M_{H_3}/M_G = 6$, $\mu < 0$ as a function of
$m_t$ as all other parameters are varied over the entire allowed parameter
space.  One sees that the lifetime peaks at $m_t \simeq 145$ GeV.  (This graph
shows again that the domain $m_o~\l~800$ GeV, $M_{H_3}/M_G < 6$ will be
accessible to Super Kamiokande.)  Fig. 5 gives a plot of the maximum value of
$\tau (p \rightarrow \bar{\nu} K)$ as a function of $m_o$ for $m_t = 150$ GeV,
$\mu > 0$, but subject to the condition $m_{\tilde{W}_1} > 100$ GeV.  One sees
that here, proton decay is accessible to ICARUS for the entire parameter space
with $M_{H_3}/M_G < 10$ (and to Super Kamiokande for $m_o~\l~950$ GeV).  Thus
for this class of models either proton decay will be seen at ICARUS or the
Wino will be seen at LEP200 (or both) and also squarks and gluinos will be
observable at the SSC or LHC.

While the $h$ Higgs boson is generally light, loop corrections can cause it to
lie beyond the planned range of LEP200.  However, by examining the full
parameter space with $m_o$, $m_{\tilde{g}} < 1$ TeV and $M_{H_3}/M_G < 10$,
one finds that if $\tau (p \rightarrow \bar{\nu} K) > 1.5 \times 10^{33}$ yr
then either $m_h < 95$ GeV or $m_{\tilde{W}_1} < 100$ GeV.  That is, if $\tau$
exceeds $1.5 \times 10^{33}$ yr (a condition that would be tested at both
Super Kamiokande or ICARUS) then either the $h$ Higgs or the $\tilde{W}_1$
(and possibly both) would be observable at LEP200.  Note also, since
$m_{\tilde{Z}_1} \simeq {1\over 2}~m_{\tilde{W}_1}$, one expects
$m_{\tilde{Z}_1}~\l~50$ GeV, for this case, and hence the $\tilde{W}_1$,
$\tilde{Z}_{1,2}$ should also be observable at the planned upgraded
Tevatron when $m_{\tilde{W}_1} < 100$ GeV (via the process [31] $p + \bar{p}
\rightarrow W^{\ast} + X \rightarrow \tilde{W}_1 + \tilde{Z}_2 + X$, with a
trileptonic plus missing $E_T$ signal).
\bigskip

\noindent
{\bf 4.} CONCLUSIONS
\medskip

An analysis of the five dimensional parameter space of $m_o$, $m_{1/2}$, $A_o$,
$\tan \beta$~and~$m_t$ for supergravity models possessing $SU(5)$-type proton
decay was carried out.  The analysis was performed under the restrictions that
(i)
the electro-weak symmetry breaking is radiative, (ii) there is no extreme fine
tuning i.e. $m_o$, $m_{\tilde{g}} \leq 1$ TeV, and (iii) $M_{H_3}/M_G \leq 10$.
It
was then shown that the intersection of the experimental limits that can be
achieved for the $\bar{\nu} K^+$ mode at SuperKamiokande and at ICARUS and the
limits on the superparticle masses achievable at LEP200 and the Tevatron can
exhaust the full parameter space of these supergravity models.  Specifically
it was shown that either the $\bar{\nu} K^+$ mode should be seen at
SuperKamiokande and ICARUS or the lighter chargino should be observable at
LEP200
provided it can achieve its optimum energy and detection efficiency.  In this
sense the proton decay experiments at SuperKamiokande and ICARUS and the LEP200
experiment are complimentary, and one needs both to check the full predictions
of
the $SU(5)$ Supergravity Model.
\bigskip

\noindent
{\bf ACKNOWLEDGEMENTS}
\medskip

We wish to thank Savas Dimopoulos for a stimulating discussion which led to
this investigation.  This research was supported in part by NSF Grant Nos.
PHY-916593 and PHY-917809.
\vfill\eject

\noindent
{\bf FIGURE CAPTIONS}
\medskip

\itemitem{Fig. 1} SUSY mass spectrum for parameters $m_o = 600$ GeV, $m_{1/2} =
53$ GeV, $A_o = 0.0$, $\tan \beta = 1.73$, $m_t = 150$ GeV and $\mu < 0$.  The
first column shows the generation 1,2 squarks and all generations of sleptons.
The
second column is the third generation of squarks.  The third column are the
charginos and neutralinos and the last column the Higgs bosons.

\itemitem{Fig. 2} One of the diagrams contributing to proton decay mode $p
\rightarrow \bar{\nu}_{\mu} + K^+$.  The Wino converts quarks into squarks,
and the baryon and lepton number violations occur at the $\tilde{H}_3$ vertex.

\itemitem{Fig. 3a} The maximum value of $\tau (p \rightarrow \bar{\nu}
K^+)$~vs~$m_o$~for~$m_t = 125$ GeV, $\mu < 0$.  The maximum is calculated by
allowing all other parameters except $m_o$ to vary over the entire allowed
parameter space.  The three curves are for $M_{H_3}/M_G = 3$, $6$, and $10$.
The
lower horizontal line is the upperbound for SuperKamiokande, i.e
SuperKamiokande
will be sensitive to lifetimes below this line.  The higher horizontal line is
for
ICARUS.

\itemitem{Fig. 3b} The same as Fig. 3a for $m_t = 150$ GeV.

\itemitem{Fig. 3c} The same as Fig. 3a for $m_t = 170$ GeV.

\itemitem{Fig. 4} The maximum value of $\tau (p \rightarrow \bar{\nu} K^+)$ vs
$m_t$ for $M_{H_3}/M_G = 6$~and~$\mu < 0$.  The solid line is for $m_o = 400$
GeV, the  dashed line for $m_o = 800$ GeV, the dot-dashed line for $m_o = 1200$
GeV.  The lower horizontal line is the bound that Super Kamiokande can detect,
and the higher horizontal line is the upperbound for ICARUS.

\itemitem{Fig. 5} Maximum value of $\tau (p \rightarrow \bar{\nu} K^+)$ vs
$m_o$
for $m_t = 150$ GeV, $\mu < 0$ when $m_{\tilde{W}_1}$ is constrained to be
greater than 100 GeV.  The solid line is for $M_{H_3}/M_G = 3$, the dashed line
is for $M_{H_3}/M_G = 6$ and the dot-dashed line is for $M_{H_3}/M_G = 10$.
The horizontal lines are as in Figs. 3 and 4.
\vfill\eject

\noindent
{\bf REFERENCES}
\medskip

\item{1.} P. Langacker, Proc. PASCOS 90-Symoposium, p. 231; Eds. P. Nath and S.
Reucroft (World Scientific, Singapore 1990); J. Ellis, S. Kelley and D. V.
Nanopoulos, Phys. Lett. \underbar{249B}, 441 (1990); \underbar{B260}, 131
(1991);
U. Amaldi, W. de Boer and H. Furstenau, Phys. Lett. \underbar{260B}, 447
(1991);
F. Anselmo, L. Cifarelli, A. Peterman and A. Zichichi, Nuov. Cim.
\underbar{115A}, 581 (1992); Nuov. Cim. \underbar{104A}, 1817 (1991).

\item{2.} A. H. Chamseddine, R. Arnowitt and P. Nath, Phys. Rev. Lett.
\underbar{49}, 970 (1982).

\item{3.} P. Nath, R. Arnowitt and A. H. Chamseddine, ``Applied N=1
Supergravity", ICTP Lecture Series Vol. I 1983 (World Scientific, Singapore,
1984); H. Nilles, Phys. Rep. \underbar{110}, 1 (1984).

\item{4.} J. Polonyi, University of Budapest Report No. KFKI-1977-93, 1977
(unpublished).

\item{5.} H. P. Nilles, Phys. Lett. \underbar{B115}, 193 (1981); S. Ferrara, L.
Girardello and H. P. Nilles, Phys. Lett. \underbar{B125}, 457 (1983); J.-P.
Derendinger, L. E. Iba$\tilde{n}$ez and H. P. Nilles, Phys. Lett.
\underbar{B155}, 65 (1985); M. Dine, R. Rohm, N. Seiberg and E. Witten, Phys.
Lett. \underbar{B156}, 56 (1985).

\item{6.} R. Arnowitt and P. Nath, Phys. Rev. Lett. \underbar{69}, 725 (1992).

\item{7.} P. Nath and R. Arnowitt, Phys. Lett. \underbar{B289}, 368 (1992).

\item{8.} J. Lopez, H. Pois, D. V. Nanopoulos and K. Yuan,
CERN-TH-6628/92-CTP-TAMU-61/92-ACT-19/92.

\item{9.} G. G. Ross and R. G. Roberts, Nucl. Phys. \underbar{B377}, 571
(1992).

\item{10.} M. Drees and M. M. Nojiri, Nucl. Phys. \underbar{B369}, 54 (1992).

\item{11.} K. Inoue, M. Kawasaki, M. Yamaguchi and T. Yanagida, Phys. Rev.
\underbar{D45}, 387 (1992); S. Kelley, J. Lopez, H. Pois, D. V. Nanopoulos and
K. Yuan, Phys. Lett. \underbar{B273}, 43 (1991).

\item{12.} P. Nath and R. Arnowitt, Phys. Lett. \underbar{B287}, 89 (1992) and
Erratum.

\item{13.} R. Arnowitt and P. Nath, TH-3056/92-CTP-TAMU-66/92 (to
be published Phys. Rev. Letts.).

\item{14.} P. Ramond, talk at Recent Advances In The Superworld, The Woodlands,
April, 1993.

\item{15.} B. Ananthanarayan, G. Lazarides and Q. Shafi, Phys. Rev.
\underbar{D44}, 1613 (1991).

\item{16.} I. Antoniadis, J. Ellis, J. S. Hagelin and D. V. Nanopoulos, Phys.
Lett. \underbar{B231}, 65 (1987); ibid, \underbar{B205} 459 (1988).

\item{17.} Particle Data Group, Phys. Rev. \underbar{D45} (1992).

\item{18.} K. Inoue et.al., Prog. Theor. Phys. \underbar{68}, 927 (1982); L.
Iba$\tilde{n}$ez and G. G. Ross, Phys. Lett. \underbar{B110} 227 (1982); L.
Alvarez-Gaum\'e, J. Polchinski and M. B. Wise, Nucl. Phys.
\underbar{B250}, 495 (1983); J. Ellis, J. Hagelin, D. V. Nanopoulos and K.
Tamvakis, Phys. Lett. \underbar{125B} 275 (1983); L. E. Iba$\tilde{n}$ez and C.
Lopez, Phys. Lett. \underbar{B128} 54 (1983); Nucl. Phys. \underbar{B233} 545
(1984); L. Iba$\tilde{n}$ez, C. Lopez and C. Munos, Nucl. Phys. \underbar{B256}
218 (1985).

\item{19.} R. Barbieri, S. Ferrara and C. A. Savoy, Phys. Lett. \underbar{119B}
343 (1982); L. Hall, J. Lykken and S. Weinberg, Phys. Rev. \underbar{D27},
2359 (1983); P. Nath, R. Arnowitt and A. H. Chamseddine, Nucl. Phys.
\underbar{B227} 121 (1983).

\item{20.} S. Coleman and E. Weinberg, Phys. Rev. \underbar{D7} 1988 (1973); S.
Weinberg, Phys. Rev. \underbar{D7} 2887 (1973).

\item{21.} G. Gamberini, G. Ridolfi and F. Zwirner, Nucl. Phys.
\underbar{B331} 331 (1990); R. Arnowitt and P. Nath, Phys. Rev. \underbar{D46}
3981 (1992).

\item{22.} H. Bethke, XXVI Conference on High Energy Physics, Dallas 1992.

\item{23.} S. Weinberg, Phys. Rev. \underbar{D26} 287 (1982); N. Sakai and  T.
Yanagida, Nucl. Phys. \underbar{B197} 533 (1982); S. Dimopoulos, S. Raby and F.
Wilczek, Phys. Lett. \underbar{112B} 133 (1982); J. Ellis, D. V. Nanopoulos and
S. Rudaz, Nucl. Phys. \underbar{B202} 43 (1982); B. A. Campbell, J. Ellis and
D. V. Nanopoulos, Phys. Lett. \underbar{141B}, 299 (1984); S. Chadha, G. D.
Coughlam, M. Daniel and G. G. Ross, Phys. Lett. \underbar{149B} 47 (1984).

\item{24.} R. Arnowitt, A. H. Chamseddine and P. Nath, Phys. Lett.
\underbar{156B} 215 (1985); P. Nath, R. Arnowitt and A. H. Chamseddine, Phys.
Rev. \underbar{32D} 2348 (1985).

\item{25.} C. D. Coughlan, G. G. Ross, R. Holman, P. Ramond, M. Ruiz-Altaba and
J. W. F. Valle, Phys. Lett. \underbar{158B} 401 (1985).

\item{26.} M. B. Gavela et.al. Nucl. Phys. \underbar{B312} 269 (1989).

\item{27.} S. Dimopoulos and H. Georgi, Nucl. Phys. \underbar{B193} 1501
(1981);
N. Sakai, Zeit. F. Phys. \underbar{C11} 153 (1981).

\item{28.} P. Langacker and N. Polonsky, Univ. Pa. Preprint 1992.

\item{29.} Y. Totsuka, Proc. of XXIV Conf. on High Energy Physics, Munich, 1988
Eds. R. Kotthaus and J. H. Kuhn (Springer Verlag, Berlin, Heidelberg, 1989).

\item{30.} ICARUS Detector Group, Int. Symposium on Nuetrino Astrophysics,
Takayama, 1992.

\item{31.} P. Nath and R. Arnowitt, \underbar{A2} 331 (1987); R. Barbieri, F.
Caravaglios, M. Frigeni and M. Mangano, Nucl. Phys. \underbar{B367} 28 (1991);
H. Baer and X. Tata, Phys. Rev. \underbar{D47}, 2739 (1993).
\vfill\eject\bye